\documentclass[12pt]{iopart}
\usepackage{cite}
\usepackage{iopams}
\usepackage{amsthm}
\usepackage{xcolor}
\providecommand{\dfrac}[2]{\displaystyle\frac{#1}{#2}}
\begin{document}
\title{Minimal sets of dequantizers and quantizers for finite-dimensional quantum systems}
\author{P. Adam$^1$, V. A. Andreev$^2$, A. Isar$^3$, M. A. Man'ko$^2$, V. I. Man'ko$^2$}
\address{$^1$Institute for Solid State Physics and Optics, Wigner Research Centre for Physics, Hungarian Academy of Sciences, H-1525 Budapest, P.O. Box 49, Hungary}
\address{$^2$P. N. Lebedev Physical Institute, Leninskii Prospect 53, Moscow 119991, Russia}
\address{$^3$National Institute of Physics and Nuclear Engineering, POB-MG6, Bucharest-Magurele, Romania}
\ead{adam.peter@wigner.mta.hu}
\begin{abstract}
  The problem of finding and characterizing minimal sets of
  dequantizers and quantizers applied in the mapping of operators onto
  functions is considered, for finite-dimensional quantum systems. The
  general properties of such sets are determined. An explicit
  description of all the minimum self-dual sets of dequantizers and
  quantizers for a qubit system is derived. The connection between
  some known sets of dequantizers and quantizers and the derived
  formulae is presented.
\end{abstract}
\submitto{\jpa}
\maketitle

\section{Introduction}\label{sec:1}
The phase-space formulation of quantum mechanics is still in the focus of research interst, as it has numerous important applications \cite{Ferrie2011, Filippov2012, Garon2015, Romero2015}.
Quasi-probability distributions such as the Wigner function \cite{WG}, the Husimi Q-function \cite{H,K} and Glauber-Sudarshan P-function \cite {G,S} describe completely the states of a quantum system and they are widely used for calculations in various physical problems \cite{CG,CG2,HOL,TT,HSW,KC}.
They have proven to be very useful in quantum optics~\cite{MW,SZ,WS}.
A probability representation with fair probability distributions defined on the phase space has also been  introduced in the literature \cite{MMT,MMT2,MMT3}.
A probability distribution called the symplectic tomogram was introduced in connection with measuring the quantum states of light by means of optical homodyne tomography \cite{Bertrand1987,Vogel1989,Smithey1993}.
The properties of this tomographic probability representation are discussed in detail in review \cite{IMM}.

In order to use quasi-probability distributions and tomograms in physical problems the operators modeling observable physical quantities have to be represented.  \cite{ST}. This representation is called the symbol of operators. The algebra of symbols corresponding all possible manipulations with operators on the Hilbert space can be constructed by applying the general star-product scheme \cite{FFL,Brif1999,MMM}. Within this formalism one can relate operators to their symbols using dequantizers and can reconstruct operators from their symbols using quantizers.
The relations between different phase-space representations can be also determined in this framework\cite{MMM,Vourdas2006,ADM1,ADM2}.

All these ideas can be extended to finite dimensional quantum systems.
Possible applications in quantum information science has generated a growing research interest aiming at the construction of discrete phase spaces and Wigner functions.
There are several ways of constructing such a phase space and the definition of a discrete Wigner function in this space is still ambiguous \cite{JSC, PT, Leonhardt1996, WKW, GHW, AV, CMM, KRS, ERL, FM1, FM2, FM3}.
The approach introduced in \cite{GHW} has proven to be well suited to study various quantum information problems \cite{Paz2005}.
In this method, an $N\times N$ phase space is defined for $N$ dimensional quantum systems, where $N$ is a power of a prime number. This is the case, e.g. for qubit systems. This phase space has the same geometric properties as those of the ordinary infinite dimensional phase space.
Wigner functions can be defined in this space using Hermitian operators connected to special mutually orthogonal sets of parallel lines called striations.
There exist $N+1$ different striations and the bases associated with them are mutually unbiased \cite{Romero2015,KRS,Bjork2007,Seyfarth2015}.
Such discrete Wigner functions have the same essential properties as their continuous counterparts. The most interesting one from the point of view of tomographic measurements is that the sum of values of a Wigner function along any line in phase space is equal to the probability of detecting the basis state associated with the line \cite{Paz2005}.

Tomographic probability distributions called spin tomograms \cite{DM,OMM,AM,AMMS}, and unitary matrix tomograms \cite{MMSV} have been also developed for finite dimensional spin systems.
The star product formalism of symbols for $N$-dimensional systems is described in detail in \cite {AAMM3}. Using this formalism the relations between tomograms and Wigner functions for one and two qubits have been determined \cite{AAM1,AAM2,AAMM3}.

In this paper we consider the problem of finding and characterizing minimal sets of quantizers and dequantizers for finite dimensional quantum systems. We determine the general properties of such sets. 
Given minimal sets of dequantizers and quantizers for a particular quantum system, any type of symbols of the operators and the quantum states consisting of minimal elements, e.g., discrete Wigner functions, can be treated in a common framework.
We find explicit expressions describing all the minimal self-dual sets of dequantizers and quantizers for a qubit system.

The paper is organized as follows.
In \sref{sec:3} we present the general formalism of mapping operators onto functions based on dequantizers and quantizers.
The general properties of minimal sets of dequantizers and quantizers for $N$ dimensional systems is described in \sref{sec:4}. In \sref{sec:5} the explicit form of all minimal self-dual sets of dequantizers and quantizers for a qubit system is found.

\section{Dequantizers and quantizers}
\label{sec:3}

In this section we summarize the general formalism of using $c$-number functions instead of operators to describe quantum systems \cite{ST,FFL,Brif1999,MMM}. Let $\hat{A}$ be a Hermitian operator acting on a Hilbert space $\mathcal{H}$ so it can be an operator describing an observable or the density operator $\hat{\rho}$ of the quantum system. Suppose we have a set of linear operators $\hat{U}(x)$ acting on $\mathcal{H}$ and labelled by the parameter $x$ that is an $n$-dimensional vector $x=(x_1,x_2,\dots,x_n)$ in the general case. One can construct a $c$-number function $f_{\hat{A}}(x)$ called the symbol of the operator $\hat{A}$ using the definition 
\begin{equation}
f_{\hat{A}}(x)=\Tr[\hat{A}\hat{U}(x)].\label{eq:fAxTr}
\end{equation}
This linear mapping of operators onto functions is invertible if there is a set of operators $\hat{D}(x)$ acting on $\mathcal{H}$ such that
\begin{equation}
\hat{A}=\int f_{\hat{A}}(x)\hat{D}(x) \rmd x.\label{eq:AintAD}
\end{equation}
The operators $\hat{U}(x)$ and $\hat{D}(x)$ are called dequantizers and quantizers, respectively.
In this formalism the operation for functions corresponding to the multiplication of $\hat{A}$ and $\hat{B}$ is called star product and defined by
\begin{equation}
f_{\hat{A}\hat{B}}(x)=f_{\hat{A}}(x)\ast f_{\hat{B}}(x)=\Tr[\hat{A}\hat{B}\hat{U}(x)].\label{eq:starproduct}
\end{equation}
Multiplying Eq.~\eref{eq:AintAD} by the operator $\hat{U}(x')$ and taking the trace we get
\begin{equation}
f_{\hat{A}}(x')=\int f_{\hat{A}}(x)\Tr[\hat{D}(x)\hat{U}(x')]\rmd x.\label{eq:fAxv}
\end{equation}

For continuous systems the operators $\hat{U}(x)$ are defined in the usual phase space with the coordinates $(q,p)$ while for discrete systems $x$ can be both discrete and continuous as in the case of spin tomograms, or it can be purely discrete as in the case of discrete Wigner functions defined e.g.\ in \cite{GHW}.
In the latter case Eqs.~\eref{eq:AintAD} and \eref{eq:fAxv} can be written as
\begin{equation}
\label{W7}
\hat{A}=\sum_{k=1}^N f_{\hat{A}}(k)\hat{D}(k)
\end{equation}
and
\begin{equation}
f_{\hat{A}}(k')=\sum_{k=1}^N f_{\hat{A}}(k)\Tr[\hat{D}(k)\hat{U}(k')],\label{eq:fAkv}
\end{equation}
respectively.

For a $d$ dimensional discrete quantum system the term minimal set of quantizers and dequantizers is introduced for sets containing $d^2$ linearly independent operators. From Eq.~\eref{eq:fAkv} it follows that the quantizer and dequantizer operators of such sets satisfy the condition
\begin{eqnarray}
\label{W9}
\Tr\big(\hat D(k)\hat U(k')\big)=\delta(k,k').
\end{eqnarray}

For some special set of dequantizers the symbols are called the Wigner function \cite{GHW}.
These dequantizers are Hermitian operators and coincide with the corresponding quantizers.
So they form a self-dual system.

\section{Minimal sets of dequantizers and quantizers}
\label{sec:4}
In this section we consider the general properties of minimal sets of quantizers and dequantizers for $N$-dimensional systems.

Let us analyse first a two-dimensional qubit system.
For this system the minimal set of dequantizers consists of four linearly independent operators $\hat U^{(k)}$ that can be represented by four matrices
\begin{equation}
\label{G1}
\hat U^{(k)}=
\pmatrix{
U_{11}^{(k)}&U_{12}^{(k)}\cr
U_{21}^{(k)}&U_{22}^{(k)}
},\qquad k=1,2,3,4.
\end{equation}

First we address the problem of determining the four corresponding quantizers
\begin{equation}
\label{G3}
\hat D^{(k)}=
\pmatrix{
D_{11}^{(k)}&D_{12}^{(k)}\cr
D_{21}^{(k)}&D_{22}^{(k)}
},\qquad k=1,2,3,4
\end{equation}
satisfying Eq.~\eref{W9} that can be written using the notations of Eqs.~\eref{G1} and \eref{G3} as 
\begin{eqnarray}
\label{eq:W9newnotation}
\Tr\big(\hat U^{(k)}\hat D^{(k')}\big)=\delta(k,k').
\end{eqnarray}
We assume that the dequantizers $U^{(k)}$ are known.

Let us introduce the operator
\begin{eqnarray}
\label{G7}
 \hat A=\pmatrix{
U_{11}^{(1)}&U_{21}^{(1)}&U_{12}^{(1)}&U_{22}^{(1)}\cr
U_{11}^{(2)}&U_{21}^{(2)}&U_{12}^{(2)}&U_{22}^{(2)}\cr
U_{11}^{(3)}&U_{21}^{(3)}&U_{12}^{(3)}&U_{22}^{(3)}\cr
U_{11}^{(4)}&U_{21}^{(4)}&U_{12}^{(4)}&U_{22}^{(4)}
},\qquad m,n=1,2,3,4
\end{eqnarray}
built up from the elements of the four dequantizer operators and the operator
\begin{eqnarray}
\label{G10}
 \hat B=\left( \begin{array}{cccc}
D_{11}^{(1)}&D_{12}^{(1)}&D_{21}^{(1)}&D_{22}^{(1)}\\
D_{11}^{(2)}&D_{12}^{(2)}&D_{21}^{(2)}&D_{22}^{(2)}\\
D_{11}^{(3)}&D_{12}^{(3)}&D_{21}^{(3)}&D_{22}^{(3)}\\
D_{11}^{(4)}&D_{12}^{(4)}&D_{21}^{(4)}&D_{22}^{(4)}
\end{array} \right)
\end{eqnarray}
containing the elements of the four quantizer operators.
It is easy to see that the equation \eref{eq:W9newnotation} is equivalent to
\begin{equation}
\hat{A}\hat{B}^T=\hat{I}\label{eq:ABeqI}
\end{equation}
As the operators $\hat{U}^{(k)}$ are linearly independent therefore the determinant of the matrix $\hat{A}$ is not equal to zero.
From Eq.~\eref{eq:ABeqI} it is clear that 
$
\det(\hat{B})=(\det(\hat{A}))^{-1}\neq 0
$ implying that the quantizers $\hat{D}^{(k)}$ are also linearly independent.
From Eq. \eref{eq:W9newnotation} or by performing the matrix product in \eref{eq:ABeqI} one can achieve four systems of equations labeled by $k'$ each of which containing four linear equations labeled by $k$ in the form
\begin{eqnarray}
\label{G6} \fl
\begin{array}{r}
U_{11}^{(k)}D_{11}^{(k')}+U_{21}^{(k)}D_{12}^{(k')}+U_{12}^{(k)}D_{21}^{(k')}+U_{22}^{(k)}D_{22}^{(k')}=\delta(k,k'),\quad k,k'=1,2,3,4.
\end{array}
\end{eqnarray}

The solution of these systems of equations can be formulated as follows:
\begin{equation}
\hat{B}=\hat{A}^{(c)}(\det(\hat{A}))^{-1},\label{eq:BeqAAc}
\end{equation}
where $\hat{A}^{(c)}$ is the cofactor matrix of $\hat{A}$.

For example the elements of the quantizer $D^{(1)}$ can be expressed as
\begin{equation}
\label{G8}
\begin{array}{ll}
D_{11}^{(1)}=A^{(c)}_{11}(\det(\hat A))^{-1},&
D_{12}^{(1)}=A^{(c)}_{12}(\det(\hat A))^{-1},\\
D_{21}^{(1)}=A^{(c)}_{13}(\det(\hat A))^{-1},&
D_{22}^{(1)}=A^{(c)}_{14}(\det(\hat A))^{-1}.
\end{array}
\end{equation}

In the following we examine the problem of finding different minimal sets
of dequantizers $\hat U^{(k)}$ and quantizers $\hat D^{(k')}$.
Given a  set of $\hat{U}^{(k)}$-s, one can derive another set $\hat{V}^{(j)}$ of dequantizers by applying a non-degenerate linear transformation $\hat{L}$:
\begin{equation}
\label{G12}
 \hat V^{(j)}=\sum_{k=1}^4L_{jk}\hat U^{(k)},\qquad j=1,2,3,4.
\end{equation}
Denoting the corresponding set of quantizers by $\hat E^{(j')}$ the novel sets will satisfy the relation
\eref{eq:W9newnotation}, that is,
\begin{equation}
\label{G13} 
\Tr(\hat V^{(j)}\hat E^{(j')})=\delta(j,j').
\end{equation}
The operators $\hat E^{(j')}$ can be viewed
as linear combinations of the previous quantizers $\hat D^{(k)}$, and thus written as
\begin{equation}
\label{G14}
 \hat E^{(j')}=\sum_{k=1}^4M_{j'k}\hat D^{(k)},\qquad j'=1,2,3,4.
\end{equation}
Substituting \eref{G12} and \eref{G14} into \eref{G13} the following matrix equation can be obtained for the four-dimensional $\hat{L}=||L_{jk}||$ and $\hat{M}=||M_{j'k}||$ matrices
\begin{equation}
\hat{L}\hat{M}^T=\hat{I}\label{eq:LMeq1}
\end{equation}
Similarly to the case of Eq.~\eref{eq:ABeqI}, this equation leads to four systems of equations each of which contains four linear equations in the form
\begin{eqnarray}
\label{G15}
\begin{array}{l}
L_{j1}M_{j'1}+L_{j2}M_{j'2}+L_{j3}M_{j'3}+L_{j4}M_{j'4}=\delta(j,j').
\end{array}
\end{eqnarray}
Solving these equations the matrix $\hat{M}$ can be written in the form
\begin{equation}
\hat{M}=\hat{L}^{(c)}(\det(\hat{L}))^{-1}\label{eq:MeqLcdetL}
\end{equation}
where $\hat{L}^{(c)}$ is the cofactor matrix of $\hat{L}$. It follows from Eq.~\eref{eq:LMeq1} that if $\hat{L}$ is non-degenerate then $\hat{M}$ is non-degenerate, too ($\det(\hat{M})=(\det(\hat{L}))^{-1}\neq 0$).

In conclusion, knowing a minimal set of dequantizers one can derive other sets by using Eq.~\eref{G12}, while Eqs.~\eref{eq:BeqAAc} and \eref{eq:MeqLcdetL} can be applied to determine the corresponding minimal sets of quantizers. 

From Eq.~\eref{G6} an additional proposition can be deduced.
In this expression the values of the matrix elements $U_{11}^{(k)}, D_{11}^{(k)}, U_{22}^{(k)}, D_{22}^{(k)}, P^{(k)}, Q^{(k)} $ are real. If the dequantizers $\hat{U}^{(k)}$ are Hermitian, that is, $U_{12}^{(k)}={U_{21}^{(k)}}^*$, in order to ensure that the expression remains real, the equality $D_{12}^{(k')}={D_{21}^{(k')}}^*$ must be satisfied. It means that the quantizers are Hermitian, too.

Let us consider the case when the dequantizers are orthogonal to each other, that is,
\begin{equation}
\Tr(\hat U^{(k)}\hat U^{(k')})=\delta(k,k').\label{eq:UUpDelta}
\end{equation}
Hence, the operators $\hat{U}^{(k)}$ form an orthogonal basis in the space of operators acting on the vectors of the Hilbert space of qubits. Comparing Eqs.~\eref{eq:W9newnotation} and \eref{eq:UUpDelta} for a fixed set of dequantizers $\hat{U}^{(k)}$ it is obvious that the choice $\hat{D}^{(k')}=\hat{U}^{(k')}$ is the solution of Eq.~\eref{eq:W9newnotation}, that is, the corresponding quantizers $\hat D^{(k)}$ coincide with dequantizers $\hat U^{(k)}$. Evidently, they are orthogonal to each other, that is,  $\Tr(\hat D^{(k)}\hat D^{(k')})=\delta(k,k')$. So the quantizers $\hat{U}^{(k)}$ and dequantizers $\hat{D}^{(k)}$ form a self-dual system.

Though all these results are formulated for a qubit, they can easily be generalized for a multi-qubit systems of higher dimensions. In $d$ dimensions the operators $\hat{A}$ and $\hat{B}$ corresponding to the operators Eqs.~\eref{G7} and \eref{G10}, respectively, can be defined by $d^2\times d^2$ matrices. The connection between $\hat{A}$ and $\hat{B}$ can be still described by Eq.~\eref{eq:BeqAAc}. The transformation rules for $\hat{U}^{(k)}$ and $\hat{D}^{(k)}$ can be easily generalized and all the previously discussed properties of these operators remain valid for higher dimensions. 
\section{Self-dual systems}\label{sec:5}

In this section we consider a qubit system and analyse in detail the case when the minimal sets of dequantizers and quantizers coincide with each other, that is, they form a self-dual system and these operators are Hermitian.

Let us take four general Hermitian operators
\begin{eqnarray}
\label{W23}
\hat U^{(k)}=\hat D^{(k)}=\pmatrix{a_k& b_k-\rmi c_k\cr b_k+\rmi c_k& d_k},\quad k=1,2,3,4,
\end{eqnarray}
where the parameters $a_k$, $b_k$, $c_k$, and $d_k$ are real.
Our aim is to find explicitly the matrix elements of the operators $\hat U^{(k)}$ so that these matrices obey Eq.~\eref{eq:UUpDelta}.

From this equation one can derive ten algebraic equations for the 16 parameters of the operators in \eref{W23}. For different matrices, that is, $k\neq k'$ we get the following six expressions:
\begin{eqnarray}
\label{W25}
\fl\begin{array}{ll}
a_1a_2+d_1d_2+2(b_1b_2+c_1c_2)=0,& a_1a_3+d_1d_3+2(b_1b_3+c_1c_3)=0,\\
a_1a_4+d_1d_4+2(b_1b_4+c_1c_4)=0,& a_2a_3+d_2d_3+2(b_2b_3+c_2c_3)=0,\\
a_2a_4+d_2d_4+2(b_2b_4+c_2c_4)=0,& a_3a_4+d_3d_4+2(b_3b_4+c_3c_4)=0,
\end{array}
\end{eqnarray}
and for $k=k'$ we have four equations:
\begin{eqnarray}
\label{eq:keqkv}
a_k^2+d_k^2+2(b_k^2+c_k^2)=1,\quad k=1,2,3,4.
\end{eqnarray}
The parameters $a_1$, $b_1$, $c_1$, $d_1$ in the operator $\hat{U}^{(1)}$  can be chosen arbitrarily, for example,
\begin{equation}
\label{W26}
a_1=1,\qquad b_1=c_1=d_1=0.
\end{equation}
Then from Eq.~\eref{W25} we get
\begin{equation}
\label{W27}
a_2=a_3=a_4=0.
\end{equation}
and the equations
\begin{equation}
\label{W28}
\begin{array}{r}
d_2d_3+2(b_2b_3+c_2c_3)=0,\\
d_2d_4+2(b_2b_4+c_2c_4)=0,\\
d_3d_4+2(b_3b_4+c_3c_4)=0.
\end{array}
\end{equation}
and Eqs.~\eref{eq:keqkv} for $k=2,3,4$, that is, six equations can be used for determining the nine remaining unknown parameters. One can obtain three different solutions of these equations depending on how many of the parameters $d_i$ are allowed to be zero. In the following the corresponding operator sets are denoted by $\hat{U}_1^{(k)}$, $\hat{U}_2^{(k)}$ and $\hat{U}_3^{(k)}$, respectively. These solutions can be characterized by the number of freely chosen parameters.

Let us assume that
\begin{equation}
\label{W29}
d_2d_3d_4\neq0.
\end{equation}
We introduce the notation
\begin{eqnarray}
\label{W32}
b_3b_4+c_3c_4=\alpha_2,\quad b_2b_4+c_2c_4=\alpha_3,\quad b_2b_3+c_2c_3=\alpha_4.
\end{eqnarray}
From Eqs.~\eref{W28} and \eref{W29} it is clear that $\alpha_i\neq 0$ and 
\begin{eqnarray}
\alpha_2\alpha_3\alpha_4=-\gamma^2<0.\label{W30}
\end{eqnarray}
In this notation the solution of the equations \eref{W28} is
\begin{equation}
\label{W31}
d_2^2=-\frac{2\alpha_3\alpha_4}{\alpha_2},\quad d_3^2=-\frac{2\alpha_2\alpha_4}{\alpha_3},\quad d_4^2=-\frac{2\alpha_3\alpha_2}{\alpha_4},
\end{equation}
and the corresponding operators $\hat{U}^{(k)}_1$ take the form 
\begin{eqnarray}
\label{E1}
\begin{array}{ll}
\hat U^{(1)}_1=\pmatrix{
1&0\cr
0&0
}&
\hat U^{(2)}_1=\pmatrix{
0&b_2-\rmi c_2\cr
b_2+\rmi c_2&\sqrt{2}\gamma/\alpha_2
},\\
\hat U^{(3)}_1=\pmatrix{
0&b_3-\rmi c_3\cr
b_3+\rmi c_3&\sqrt{2}\gamma/\alpha_3
},&
\hat U^{(4)}_1=\pmatrix{
0&b_4-\rmi c_4\\
b_4+\rmi c_4&\sqrt{2}\gamma/\alpha_4
}.
\end{array}
\end{eqnarray}
For the normalization equations \eref{eq:keqkv} we get
\begin{eqnarray}
2\gamma^2/\alpha_k^2+2b_k^2+2c_k^2=1,\quad k=2,3,4.\label{eq:case1conds}
\end{eqnarray}
These equations contain six parameters, hence, three of them can be chosen freely within certain constraints ensuring that the equations have real solutions satisfying the condition in Eq.~\eref{W30}.\\
The general form of the solutions of Eqs.~\eref{eq:case1conds} is complicated and depends on the three chosen independent parameters. Without presenting here these expressions explicitly we present one set of dequantizer operators corresponding to the general form in Eq.~\eref{E1}:
\begin{eqnarray}
\label{eq:case1expl}
\begin{array}{ll}
\hat U_1^{(1)}=\pmatrix{
1&0\cr
0&0
}&
\hat U_1^{(2)}=\dfrac{1}{4}\pmatrix{
0&2-\rmi \cr
2+\rmi &-\sqrt{6}
},\\
\hat U_1^{(3)}=\dfrac{1}{4}\pmatrix{
0&2+\rmi\cr
2-\rmi&\sqrt{6}
},&
\hat U_1^{(4)}=\dfrac{1}{4}\pmatrix{
0&-\rmi \sqrt{6}\\
\rmi \sqrt{6}&2
}.
\end{array}
\end{eqnarray}

Next, let us apply the conditions
\begin{equation}
\label{W33}
d_2=0,\qquad d_3d_4\neq0.
\end{equation}
In this case the equation \eref{W28} reads
\begin{equation}
\label{W34}
b_2b_3+c_2c_3=0,\quad b_2b_4+c_2c_4=0,\quad d_3d_4+2(b_3b_4+c_3c_4)=0.
\end{equation}
Let us choose the parameters $b_4,c_2,c_3,c_4,d_4$ to be the independent ones in these equations. Then the three remaining parameters can be expressed as
\begin{equation}
\label{W36}
b_2=-\frac{c_2c_4}{b_4}, \qquad b_3=\frac{b_4c_3}{c_4}, \qquad d_3=-\frac{2c_3}{c_4d_4}(b_4^2+c_4^2).
\end{equation}
In this case the matrices of the dequantizer operators read
\begin{eqnarray}\fl
\begin{array}{ll}
\label{E2var}
\label{E2}
\hat U_2^{(1)}=\pmatrix{1&0\cr0&0},&
\hat U_2^{(2)}=-\dfrac{c_2}{b_4}\pmatrix{0&c_4+\rmi b_4\cr c_4-\rmi b_4&0},\\
\hat U_2^{(3)}=\dfrac{c_3}{c_4}\pmatrix{0&b_4-\rmi c_4\cr b_4+\rmi c_4&-\frac{2}{d_4}(b_4^2+c_4^2)},&
\hat U_2^{(4)}=\pmatrix{0&b_4-\rmi c_4\cr b_4+\rmi c_4&d_4}.
\end{array}
\end{eqnarray}
For these operators the normalization conditions take the form
\begin{eqnarray}
\frac{2c_2^2c_4^2}{b_4^2}+2c_2^2&=&1,\label{case2norm1}\\
\frac{8b_4^2c_3^2+4c_3^2c_4^2}{d_4^2}+\frac{4b_4^4c_3^2}{c_4^2d_4^2}+\frac{2b_4^2c_3^2}{c_4^2}+2c_3^2&=&1,\label{case2norm2}\\
d_4^2+2c_4^2+2b_4^2&=&1,\label{case2norm3}
\end{eqnarray}
from which the following expressions can be derived:
\begin{eqnarray}
b_4&=&\pm c_2\sqrt{\frac{2c_3^2+2c_2^2-1}{2c_2^2-1}},\label{eq:case2b4}\\
c_4&=&\pm \sqrt{\frac{-2c_3^2-2c_2^2+1}{2}},\label{eq:case2c4}\\
d_4&=&\pm \frac{\sqrt{2}\rmi c_3}{\sqrt{2c_2^2-1}}.\label{eq:case2d4}
\end{eqnarray}
Recall that all the parameters are chosen to be real, leading to a restriction in the choice of $c_2$  and $c_3$ in Eqs.~\eref{eq:case2b4}--\eref{eq:case2d4} ($|c_2|\leq1/\sqrt{2}$ and $|c_3|\leq1/\sqrt{2}$). As an example, we choose $c_2=c_3=\dfrac{1}{2\sqrt{2}}$. Substituting these values to formulas \eref{E2var} the following operators can be derived:
\begin{eqnarray}\fl
\label{eq:case2expl}
\begin{array}{ll}
\hat U_2^{(1)}=\pmatrix{1&0\cr0&0},&
\hat U_2^{(2)}=-\dfrac{1}{2\sqrt{2}}\pmatrix{0&\sqrt{3}+\rmi\cr\sqrt{3}-\rmi&0},\\
\hat U_2^{(3)}=\dfrac{1}{\sqrt{6}}\pmatrix{0&\dfrac{1-\rmi\sqrt{3}}{2}\cr\dfrac{1+\rmi\sqrt{3}}{2}&-2},&
\hat U_2^{(4)}=\dfrac{1}{\sqrt{3}}\pmatrix{0&\dfrac{1-\rmi\sqrt{3}}{2}\cr\dfrac{1+\rmi\sqrt{3}}{2}&1}.
\end{array}
\end{eqnarray}

Finally, let us assume that
\begin{equation}
\label{W37}
d_2=d_3=0,\qquad d_4\neq0.
\end{equation}
In this case the system \eref{W28} reads
\begin{equation}
\label{W38}
b_2b_3+c_2c_3=0,\quad b_2b_4+c_2c_4=0,\quad b_3b_4+c_3c_4=0.
\end{equation}
Assuming $b_3,c_2,c_3,d_4$ to be the independent parameters, the solutions of these equations are
\begin{eqnarray}
\label{W39}
b_2=-\frac{c_2c_3}{b_3},\quad b_4=0,\quad c_4=0,
\end{eqnarray}
leading to the following dequantizers:
\begin{eqnarray}\fl
\begin{array}{ll}
\label{E3}
\hat U_3^{(1)}=\pmatrix{1&0\cr0&0},&
\hat U_3^{(2)}=-\dfrac{c_2}{b_3}\pmatrix{0&c_3+\rmi b_3\cr c_3-\rmi b_3&0},\\[5mm]
\hat U_3^{(3)}=\pmatrix{0&b_3-\rmi c_3\cr b_3+\rmi c_3&0},&
\hat U_3^{(4)}=\pmatrix{0&0\cr0&d_4}.
\end{array}
\end{eqnarray}
The equations describing the normalization conditions read
\begin{eqnarray}
\begin{array}{rcl}
\dfrac{2c_2^2c_3^2}{b_3^2}+2c_2^2&=&1,\\
2c_3^2+2b_3^2&=&1,\\
d_4^2&=&1,
\end{array}
\end{eqnarray}
resulting in the expressions
\begin{eqnarray}
\label{eq:case2cond}
\begin{array}{rcl}
c_2&=&\pm b_3,\\
c_3&=&\pm \dfrac{\sqrt{1-2b_3^2}}{\sqrt{2}},\\
d_4&=&\pm 1.
\end{array}
\end{eqnarray}
From equations \eref{eq:case2cond} it is clear that in the choice of $b_3$ the condition $|b_3|\leq \dfrac{1}{\sqrt{2}}$ must be satisfied.
As an example, by choosing $b_3=\dfrac{1}{2}$ we can get $c_2=c_3=\dfrac{1}{2}$, and $d_4=1$, leading to the dequantizer operators
\begin{eqnarray}
\label{eq:case3expl}
\begin{array}{ll}
\hat U_3^{(1)}=\pmatrix{1&0\cr0&0},&
\hat U_3^{(2)}=-\dfrac12\pmatrix{0&1+\rmi\cr1-\rmi&0},\\[5mm]
\hat U_3^{(3)}=\dfrac12\pmatrix{0&1-\rmi\cr1+\rmi&0},&
\hat U_3^{(4)}=\pmatrix{0&0\cr0&1}.
\end{array}
\end{eqnarray}

Here we should point out that the three different sets of dequantizers and quantizers presented in Eqs.~\eref{E1}, \eref{E2}, and \eref{E3} contain three, two, and one parameters that can be chosen freely, respectively, taking into account the normalization conditions in these expressions. In the following we consider the connection of them with some of the minimal sets presented in the literature thus far.

In \cite{AAM1,AAM2} the following self-dual minimal system of dequantizers are considered for deriving a discrete Wigner function
\begin{eqnarray}
\label{W40}\fl
\begin{array}{ll}
\hat{V}^{(1)}=\dfrac1{\sqrt2}\pmatrix{1&\frac{1}{2}(1-\rmi)\cr\frac{1}{2}(1+\rmi)&0},&
\hat{V}^{(2)}=\dfrac1{\sqrt2}\pmatrix{1&\frac{1}{2}(-1+\rmi)\cr\frac{1}{2}(-1-\rmi)&0},\\[5mm]
\hat{V}^{(3)}=\dfrac1{\sqrt2}\pmatrix{0&\frac{1}{2}(1+\rmi)\cr\frac{1}{2}(1-\rmi)&1},&
\hat{V}^{(4)}=\dfrac1{\sqrt2}\pmatrix{0&\frac{1}{2}(-1-\rmi)\cr\frac{1}{2}(-1+\rmi)&1}.
\end{array}
\end{eqnarray}
The matrices \eref{W40} can be presented as linear combinations of the constructed sets of dequantizers.
In the following we show how the matrices $\hat{V}^{(j)}$ in \eref{W40} can be constructed from any sets of matrices $U^{(k)}_i$ ($i=1,2,3$) presented in Eqs.~\eref{eq:case1expl}, \eref{eq:case2expl}, and \eref{eq:case3expl}, respectively, and we determine the corresponding linear transformations $L^{(i)}$.

Let us apply formula \eref{G12} for this case
\begin{equation}
\label{eq:G12}
 \hat V^{(j)}=\sum_{k=1}^4L^{(i)}_{jk}\hat U^{(k)}_i,\qquad j=1,2,3,4.
\end{equation}
Multiplying both sides by $\hat{U}^{(k')}_i$ and taking the trace of the expressions we get
\begin{equation}
 \Tr(\hat V^{(j)}\hat{U}^{(k')}_i)=\Tr\left(\sum_{k=1}^4L^{(i)}_{jk}\hat U^{(k)}_i\hat{U}^{(k')}_i\right)=\sum_{k=1}^4L^{(i)}_{jk}\Tr(\hat U^{(k)}_i\hat{U}^{(k')}_i).
\end{equation}
Using Eq.~\eref{eq:UUpDelta} the expression on the right hand side simplifies to $L^{(i)}_{jk'}$, therefore
\begin{equation}
 \Tr(\hat V^{(j)}\hat{U}^{(k')}_i)=L^{(i)}_{jk'}.\label{eq:findLfromVandU}
\end{equation} 
Applying this formula the linear transformation $L^{(1)}$ connecting $\hat{V}^{(j)}$ and $\hat{U}^{(k)}_1$ takes the form
\begin{equation}
L^{(1)}=\frac{1}{2^{5/2}}\pmatrix{
4 & 3 & 1 & \sqrt{6} \cr
4 & -3 & -1 & -\sqrt{6} \cr
0 & 1-\sqrt{6} & \sqrt{6}+3 & 2-\sqrt{6}\cr
0 & -\sqrt{6}-1 & \sqrt{6}-3 & \sqrt{6}+2
},
\end{equation}
while for $L^{(2)}$ and $L^{(3)}$ transforming $\hat{U}^{(k)}_2$ and $\hat{U}^{(k)}_3$ to $\hat{V}^{(j)}$ we obtain
\begin{equation}
L^{(2)}=\frac{1}{4\sqrt{3}}\pmatrix{
2\sqrt{6} & \sqrt{3}-3 & \sqrt{3}+1 & \sqrt{2}\left(\sqrt{3}+1\right)
\cr
2\sqrt{6} & 3-\sqrt{3} & -\sqrt{3}-1 & -\sqrt{2}\left(\sqrt{3}+1\right)
\cr
0 & -\sqrt{3}-3 & -\sqrt{3}-3 & \sqrt{2}\left(3-\sqrt{3}\right)
\cr
0 & +\sqrt{3}+3 & \sqrt{3}-5 & \sqrt{2}\left(\sqrt{3}+1\right)}
\end{equation}
and
\begin{equation}
L^{(3)}=\frac{1}{\sqrt{2}}
\pmatrix{
1 & 0 & 1 & 0\cr
1 & 0 & -1 & 0\cr
0 & -1 & 0 & 1\cr
0 & 1 & 0 & 1
}.
\end{equation}

Another example for a set of dequantizers are the ones presented in Ref.~\cite{ERL} for deriving a discrete Wigner function where the operators $V^{(k)}$ were denoted by $D_{-+}$, $D_{++}$, $D_{--}$, and $D_{+-}$.
The explicit form of these operators are
\begin{eqnarray}
\label{LivineDs}\fl
\begin{array}{ll}
\hat{V}^{(1)}=\dfrac{1}{4}\pmatrix{
0&1+\rmi\cr
1-\rmi&2
},&
\hat{V}^{(2)}=\dfrac{1}{4}\pmatrix{2&1-\rmi\cr 1+\rmi&0},\\[5mm]
\hat{V}^{(3)}=\dfrac{1}{4}\pmatrix{0&-1-\rmi\cr -1+\rmi&2},&
\hat{V}^{(4)}=\dfrac{1}{4}\pmatrix{2&-1+\rmi\cr -1-\rmi&0}.
\end{array}
\end{eqnarray}
Applying Eq.~\eref{eq:findLfromVandU} for these operators the corresponding linear transformation $L^{(i)}$ can be found for any of the above sets $\hat{U}^{(k)}_i$. As an example, the linear transformation $L^{(3)}$ for the set of dequantizers presented in \eref{eq:case3expl} takes the form
\begin{equation}
L^{(3)}=\frac{1}{2}
\pmatrix{
0 & -1 & 0 & 1\cr
1 & 0 & 1 & 0\cr
0 & 1 & 0 & 1\cr
1 & 0 & -1 & 0
}
\end{equation}
From these results one can conclude that any minimal sets of dequantizers used for defining discrete Wigner functions can be derived from any type of the minimal sets presented in Eqs.~\eref{E1}, \eref{E2}, and \eref{E3} using the described procedure. An interesting aspect of our results is that we found as many different types of minimal sets, i.e., three, as the number of mutual unbiased bases in this space. Recall that the different types of discrete Wigner functions that can be derived according to the approach introduced in \cite{GHW} are associated with the mutually unbiased bases.

The generalization of the presented method of finding minimal sets of dequantizers and quantizers for higher dimensions is not obvious owing to the large number of parameters. In $d$ dimensions the number of operators forming the minimal set is $d^2$ containing $d^4$ parameters, while the number of equations following from the orthogonality and normalization conditions is $d^2(d^2+1)/2$.
Nonetheless, one can construct the minimal sets of dequantizers and quantizers for an $N$-qubit system using the tensor product of one-qubit dequantizers and quantizers.

\section{Conclusion}
\label{sec:concl}
We analysed the general properties of minimal sets of dequantizers and quantizers for finite-dimensional quantum systems. 
We developed a general approach for deriving the corresponding quantizers assuming that a minimal set of dequantizers is known, and we have descrbed the connection between different minimal sets.
We have derived explicit expressions describing all minimal sets of dequantizers and quantizers for a qubit. We have shown explicitly how some known sets of dequantizers and quantizers used in certain problems in the literature can be derived from these formulae.

\ack

This work was performed within the framework of the collaboration between the Hungarian Academy of Sciences and the Russian Academy of Sciences on the Problem ``Quantum Correlations, Decoherence in the Electromagnetic Field Interaction with Matter and Tomographic Approach to Signal Analysis'' and between the Romanian Academy of Sciences and the Russian Academy of Sciences on the Problem ``Fundamental Aspects of Quantum Optics and Quantum Correlations in Information Theory.'' A.I. acknowledges the financial support from the Romanian Ministry of Education and Research under Project~CNCS-UEFISCDI PN-II-ID-PCE-2011-3-0083.

\section*{References}  
\providecommand{\newblock}{}

\end{document}